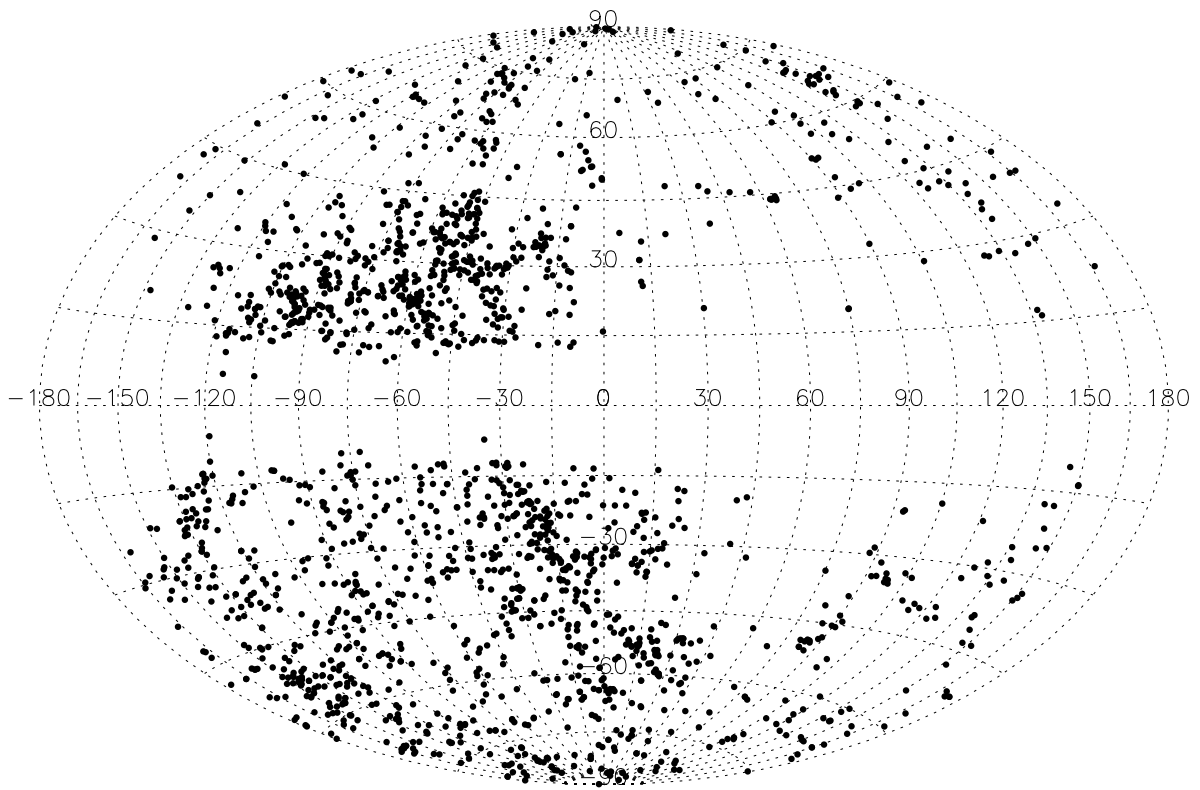

Figure 1:

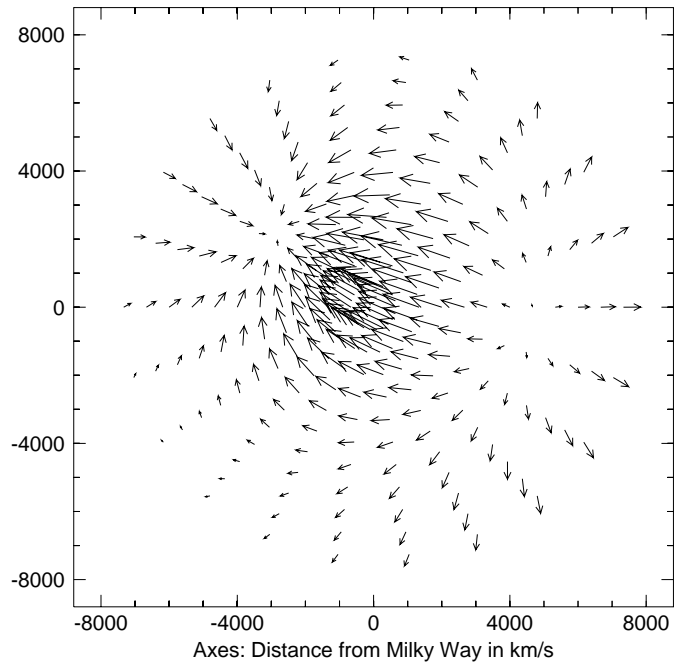

Figure 2:

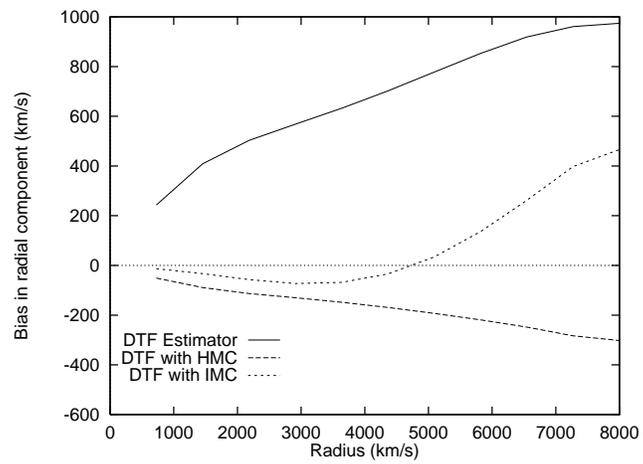

Figure 3:

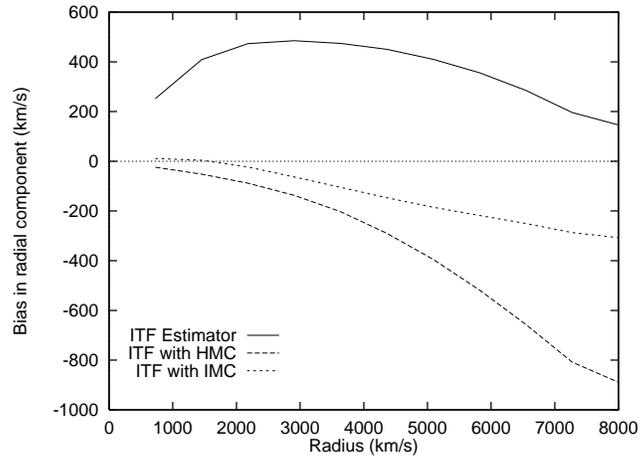

Figure 4:

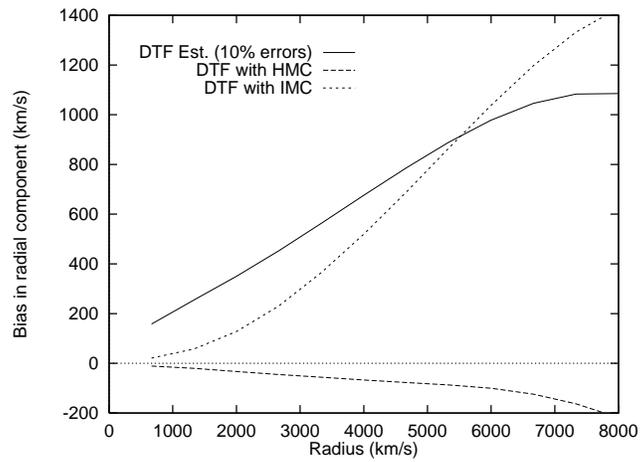

Figure 5:

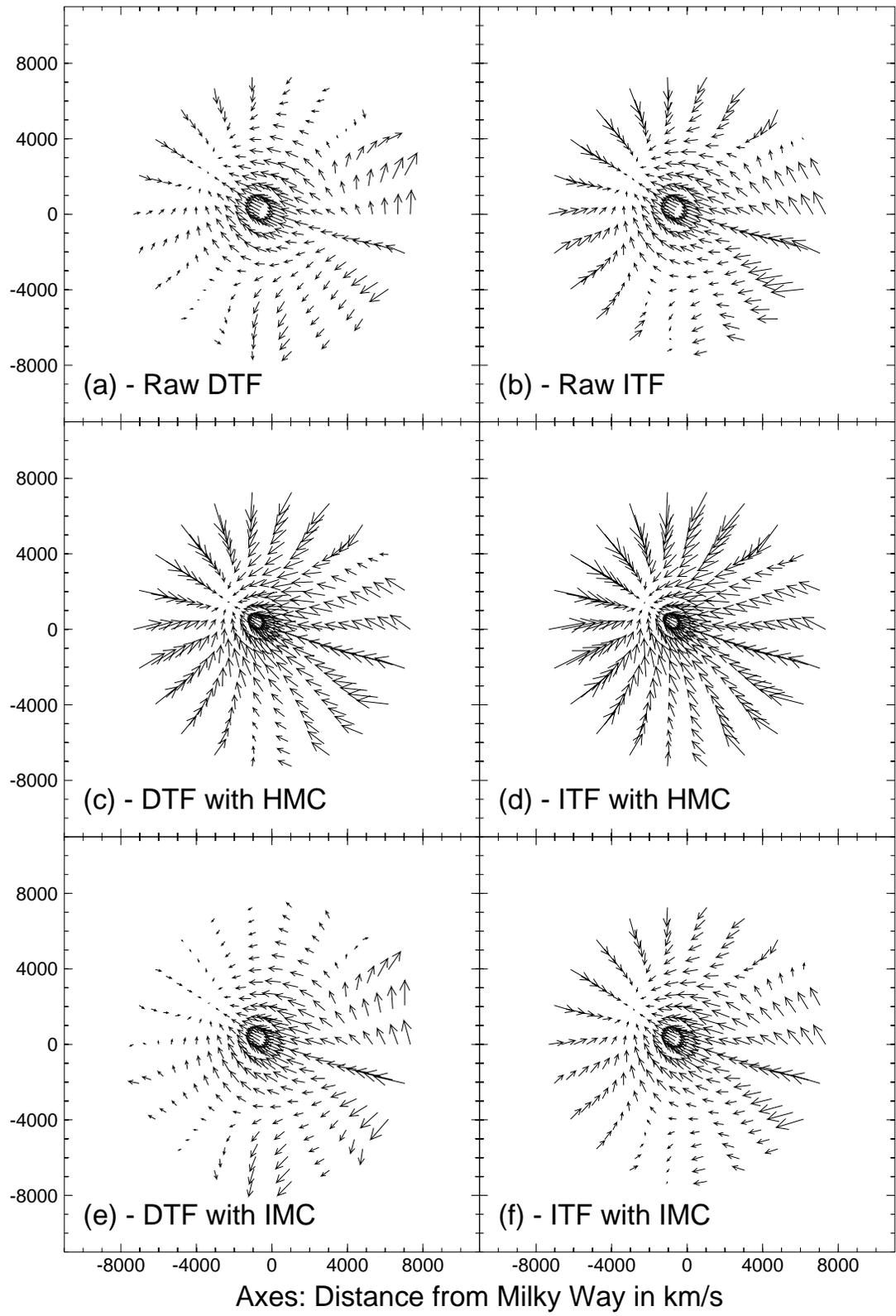

Axes: Distance from Milky Way in km/s

Figure 6:

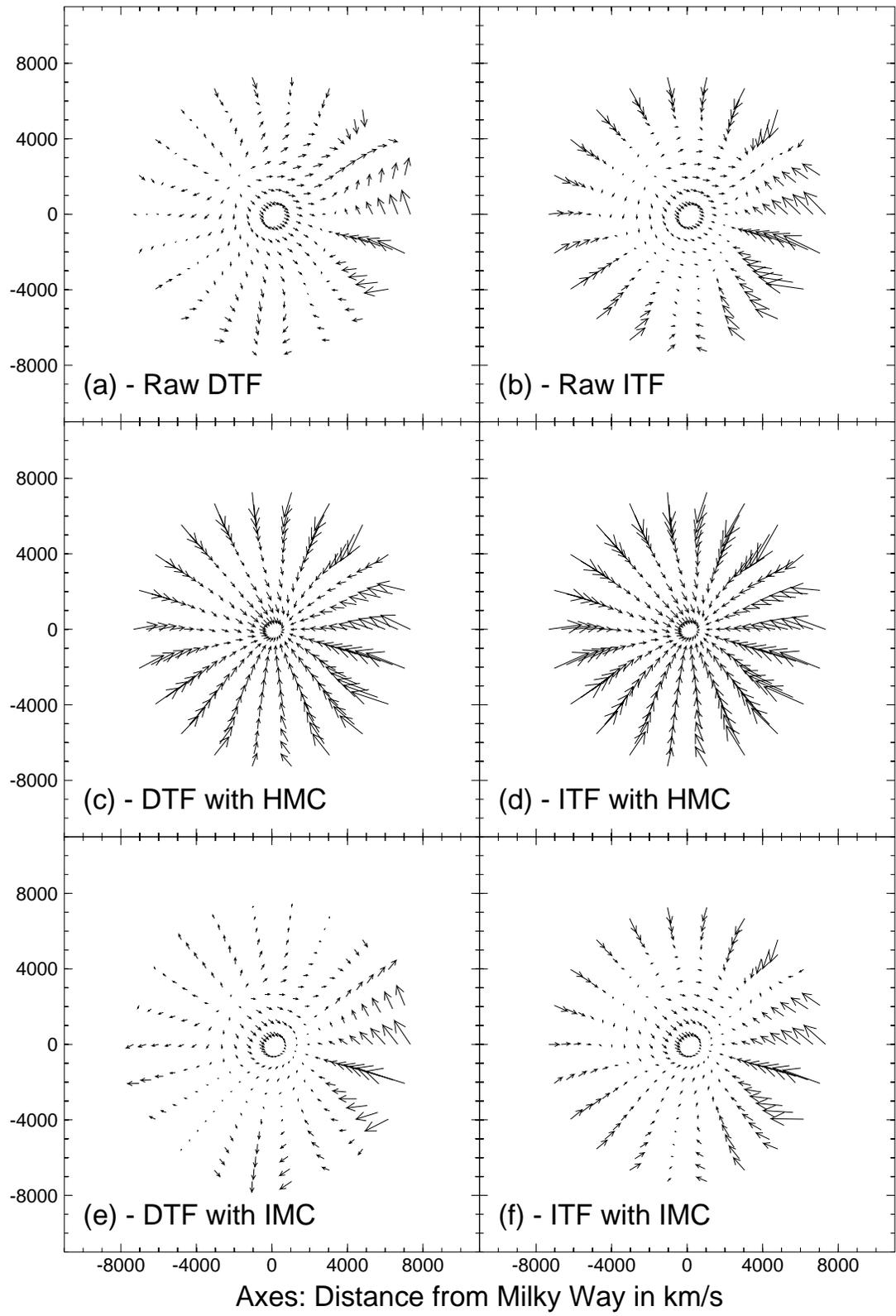

Axes: Distance from Milky Way in km/s

Figure 7:

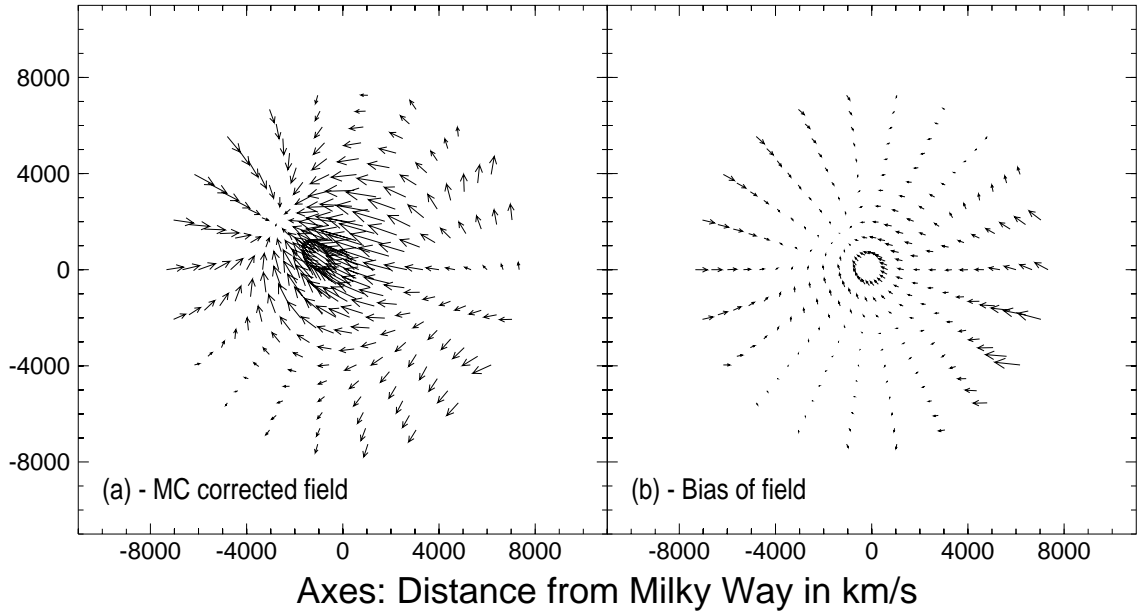

Figure 8:

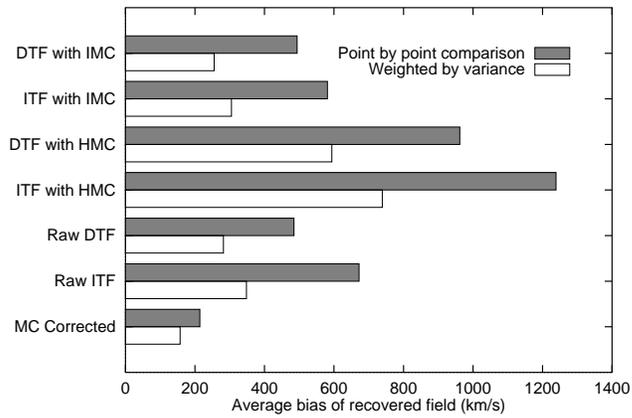

Figure 9:

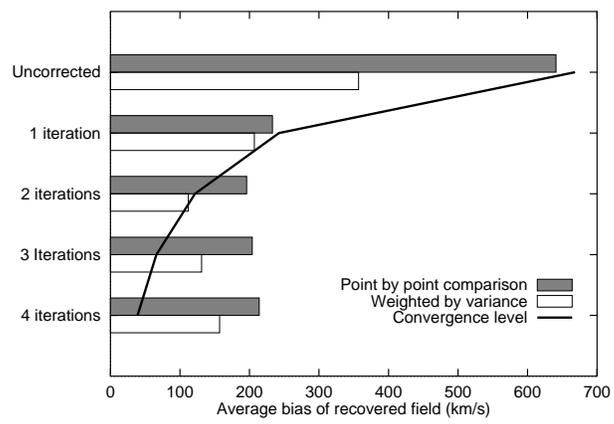

Figure 10:

# Bias minimisation in Potent

A. Newsam[1], J.F.L. Simmons[1], and M.A. Hendry[2]

[1] Department of Physics and Astronomy, University of Glasgow, Glasgow, UK
[2] Department of Astronomy, University of Sussex, Falmer, Brighton, UK

**Abstract.** Much current interest in cosmology has revolved around the peculiar velocity field of galaxies. Many methods have been proposed to derive this, probably the most prominent being POTENT (Dekel et al. 1990). However, this is prone to many forms of systematic error or bias. Here, we examine some of the methods currently used to minimise these biases, particularly Malmquist corrections, and highlight some of their pitfalls. As an alternative we propose an iterative scheme based on Monte Carlo simulations of POTENT that shows noticeable improvement over the other methods considered.

**Key words:** cosmology – galaxies: redshifts of – numerical methods – radial velocities – universe: structure of

## 1. Introduction

Recently cosmologists have paid a great deal of attention to the problem of constructing the peculiar velocity field of galaxies. The main importance of such velocity field reconstructions lies in their implications for the large scale distribution of matter: any systematic deviations of the velocity field from quiet Hubble flow indicate density inhomogeneities, the measurement of which can place constraints upon $\Omega$. Linear perturbation theory yields the well known relationship

$$\nabla \cdot \vec{v} = -f(\Omega)\delta \qquad (1)$$

where $f(\Omega) = \Omega^{0.6}$

and $\delta$ is the density contrast.

The essential idea of POTENT (Bertschinger and Dekel 1989; Bertschinger et al., 1990; Dekel et al., 1990 hereafter DBF; Dekel et al., 1992) is to derive this underlying peculiar velocity field directly from measurements of galaxy redshifts. The fundamental assumption is that the peculiar velocity field can be represented by a gradient of a scalar potential function. It then follows that this potential can be derived by taking the line integral of the radial peculiar velocity components $u$ along a radial path.

Of course, to do this, we need to know these components at all points along the path and even with the recent increase in suitable galaxy samples (for example Mathewson et al., 1992) this can only be done by smoothing the data from

*Send offprint requests to*: A. Newsam

large volumes (of the order of thousands of km s$^{-1}$ in diameter). In addition, there are errors of up to 20% in the determination of distances to galaxies and, hence, highly significant errors in $\hat{\mathbf{u}}$[1]

$$\hat{\mathbf{u}} = cz - H_0 \hat{\mathbf{r}} \qquad (2)$$

where $z$ is the redshift and $\hat{\mathbf{r}}$ is the estimated distance to a galaxy.

The combination of these two problems can produce large systematic errors or biases in POTENT which, if not corrected adequately, can lead to very convincing but totally spurious coherent velocity fields. The aim of this paper is to examine some of the methods used to date to compensate or correct for these biases, and then to present a new method that attempts to correct for all these systematic errors simultaneously.

Current methods involve applying various so-called 'Malmquist corrections', which attempt to remove systematic errors arising from the smoothing – where galaxies are grouped together according to their estimated distance. In Sects. 3 and 4 we discuss some important statistical properties of galaxy distance estimators – summarising results from our earlier papers and other relevant references – particularly concerning the valid definition and use in POTENT of Homogeneous Malmquist corrections (Lynden-Bell et al., 1988) and General (or Inhomogeneous) Malmquist corrections (Landy and Szalay, 1992). Some numerical tests to demonstrate the effects of using different combinations of estimators and corrections on both simplified and realistic data sets are then given in Sect. 5.

The new method we introduce in Sect. 6 is an iterative Monte Carlo technique that tries to remove all biases simultaneously by slowly adjusting some estimate of the field until, when passed through POTENT, it matches the result from the actual data. Results of some realistic tests are given in Sect. 7 and these compared to the results of Sect. 5.

## 2. How POTENT Works

Full descriptions of the POTENT method are given by its originators in DBF, but it is convenient to summarise here the basic ideas of the method so that we can subject some of the assumptions to closer scrutiny.

---

[1] Please note that for this paper, we will be using the standard statistical notation for estimators etc. This means that an estimate of some quantity $x$ will have a 'hat', $\hat{\mathbf{x}}$ and statistical variables will be in bold face $\mathbf{x}$. Also, $E(\mathbf{x_1}|\mathbf{x_2})$ will be used to denote the expectation value of the quantity $\mathbf{x_1}$ given $x_2$.

The key idea of POTENT is to write

$$\vec{v}(r, \theta, \phi) = -\nabla \Phi(r, \theta, \phi) \tag{3}$$

and hence obtain $\Phi(r, \theta, \phi)$ from a suitable line integral. Taking a radial path will involve only the radial components of the velocity fields, and hence

$$\Phi(r, \theta, \phi) = -\int_0^r u \, dr \tag{4}$$

where $u$ is given by the redshifts similarly to Eq. (2).

The problem in carrying out this radial line integral is that

1. the radial component of the peculiar velocity can only be obtained at those points where galaxies are found and
2. these peculiar velocities are only estimates, and rely on the estimation of the distances to galaxies as in Eq. (2).

To cope with sparseness DBF use tensor window functions. This method obtains a smoothed peculiar velocity field, $\vec{v}$, at every spatial point, by best fitting to the estimated radial components of the peculiar velocity $\hat{u}$. This is done by minimising

$$\sum_{i=1}^{N} (\hat{\vec{v}} \cdot \vec{e}_r(\hat{\vec{r}}_i) - \hat{u}_i)^2 \, W(\vec{r}, \hat{\vec{r}}_i) \tag{5}$$

with respect to $\vec{v}(\vec{r})$, where $W(\vec{r}, \vec{r}_i)$ is a chosen window function, $\hat{\vec{r}}_i$ is the estimated position vector of the $i^{\text{th}}$ galaxy, and there are $N$ galaxies is the catalogue. $\vec{e}_r(\vec{r}_i)$ is the unit vector in the radial direction of the $i^{\text{th}}$ galaxy at position $\vec{r}_i$. (N.B. $\vec{r}_i$ and $\hat{\vec{r}}_i$ will be in the same direction.)

There are several crucial points to observe in this procedure.

1. Even in the absence of distance errors the smoothed peculiar velocity field will not be exact. DBF consider the case where an input smooth peculiar velocity field is sampled at various points corresponding to galaxies, and reconstructed using procedure given by Eq. (5). This rederived field will be subject to 'sampling gradient bias', which will be particularly acute where the galaxies are sparse. A good choice of window function will help to minimise this effect, but for spatially inhomogeneous samples the effect cannot be removed everywhere.
2. If noise is introduced into the smoothed input field, and/or distance estimates are subject to errors, the mean retrieved field obtained through smoothing will in general not be equal to the input smooth field. In other words the retrieved smoothed field will be biased. DBF call this Malmquist bias, in distinction to the sampling gradient bias, and have attempted to remove it by applying homogeneous Malmquist corrections (Lynden-Bell et al. 1988 hereafter LB) to the distance estimates.

## 3. Bias of Distance Estimators

The past decade has witnessed rapid development in the use of redshift – independent galaxy distance indicators, such as the Tully–Fisher (TF) relation, which rely upon the existence of an observable – denoted generically by P – that correlates strongly with e.g. the absolute magnitude of a galaxy. Thus by measuring P, and then by measuring the *apparent* magnitude, a distance estimate to the galaxy may be determined.

It is instructive to summarise briefly some of the statistical properties of such 'TF–type' relations, although we do no more than recapitulate those points which are relevant to a discussion of distance indicators as they are used in POTENT. For a more detailed discussion see e.g. Hendry (1992), Hendry et al. (1994a), Hendry and Simmons, (1994, hereafter HS) and references therein.

The TF relation is usually fitted by some kind of linear regression. It is easy to see that the estimator, $\hat{\omega}$, of log distance which corresponds to a linear TF relation takes the form

$$\hat{\omega} = 0.2(\mathbf{m} - \mathbf{aP} - \mathbf{b} - 25) \tag{6}$$

where $a$ and $b$ are constants. (c.f. Eq. (11) of HS). In general $\hat{\omega}$ is a biased estimator: i.e.

$$E(\hat{\omega}|\omega_0) \neq \omega_0 \tag{7}$$

where the expectation value is taken over the observable population of galaxies at fixed true log distance, $\omega_0$, having taken selection into account. In HS, however, we proved that if galaxies are selected only by their apparent magnitude then $\hat{\omega}$ is unbiased for all $\omega_0$, provided that $a$ and $b$ in Eq. 6 are derived from a regression of $\mathbf{P}$ on $\mathbf{M}$, the so–called 'inverse Tully Fisher' (ITF) relation. The estimator, $\hat{\omega}_{\text{DTF}}$ corresponding to a 'direct' regression of $\mathbf{M}$ on $\mathbf{P}$ is, on the other hand, biased for all $\omega_0$. The unbiased property of the ITF relation was first pointed out by Schechter 1980, and has been generally recognised in the literature, although few discussions have approached the subject in a fully rigorous manner.

More generally, in Sect. (2.4) of HS we proved that the distribution of $\hat{\omega}_{\text{ITF}}$, conditional upon $\omega_0$, is gaussian for all $\omega_0$ – provided only that the conditional distribution of $\mathbf{P}$ given $\mathbf{M}$ is gaussian, and that $E(\mathbf{P}|\mathbf{M})$ is a linear function of $\mathbf{M}$. The distribution of $\hat{\omega}_{\text{ITF}}$, however – or indeed the distribution of the estimator derived from *any* regression other than $\mathbf{P}$ on $\mathbf{M}$ – is in general non–gaussian and biased for all $\omega_0$. Thus, if galaxies are subject to selection only on magnitudes, then one must use the ITF estimator in order to obtain normally distributed, unbiased estimates of $\log r$.

Several authors (c.f. LB, Landy and Szalay (1992) hereafter LS) take a different approach to the problem of defining unbiased distance estimators, leading to their calculation of Malmquist Corrections. In HS the differences between these two approaches are discussed in detail, and the use of Malmquist Corrections is shown to reflect a fundamentally Bayesian view of the problem. It is this latter, Bayesian, approach which has been adopted in the treatment of distance errors in POTENT (DBF, Dekel et al. 1992). In the next section we will consider briefly why this has been the case, and recall from HS some important results concerning the assumptions upon which the calculation of Malmquist corrections has been based.

## 4. The Treatment of Distance Estimators in POTENT

Whether or not a distance estimator is biased is not the crucial question when attempting to correct for bias in POTENT. What is important is to construct an unbiased smoothed peculiar velocity field. POTENT attempts to construct an unbiased peculiar velocity field in the following sense:

(taken to be potential) and effective density distribution that is determined by some selection function and underlying density distribution of galaxies.

1. This field is sampled at $n$ points and the galaxies taken to be at these points corresponding to the actual distances $(r_{10}, r_{20}, r_{30}, ..., r_{n0})$
2. Errors are added to these distances. A smoothed initial radial peculiar velocity field is derived using the tensor window function.
3. Hence, one obtains a potential velocity field by radial integration.

If this smoothed recovered potential velocity field is the same as the input potential velocity field when it is averaged over all realisations of $(r_{10}, r_{20}, r_{30}..r_{n0})$ and of the distance errors, it is unbiased.

In the Appendix of DBF an attempt is made to prove that if one applies a homogeneous Malmquist correction to the raw distance estimates then one does obtain a peculiar velocity field which is almost unbiased. Essentially their analysis proceeds by expressing the bias of the recovered velocity field as a function of the errors, $\epsilon_i$, in the galaxy distance estimates, and depends upon making several Taylor expansions in $\epsilon_i$ and discarding terms of order 3 and above. If the distance errors are large, as they will be at large true distances, this procedure will break down. DBF employ Monte Carlo simulations to back up their analytic treatment.

It is instructive to examine why the application of Malmquist corrections appears to work for the POTENT analysis. In this respect, the important factor is the window function. In interpolating a peculiar velocity from galaxies appearing in the catalogue to a given spatial point with radial coordinate $s$, the *essential* effect of the window function is to pick out the galaxy whose *estimated* position is nearest to the prescribed point. This galaxy's actual distance could be radically different, and will depend on the true spatial distribution of galaxies. By requiring that on average the actual radial coordinate of the galaxy deemed to be closest to the grid point equals $s$ one would ensure also that on average the correct peculiar velocity would be ascribed to $s$. Expressed mathematically we require

$$E(\mathbf{r}_0|\hat{\mathbf{r}} = \mathbf{s}) = \mathbf{s} \qquad (8)$$

which is precisely the condition defining Malmquist–corrected distance estimators – c.f. Eq. (7) of HS, Eq. (9) of LS.

Of course this will strictly only be valid if galaxies are not too sparse, and if the gradient of velocity field is not too large, or the effective radius of the window function is not too wide.

*4.1. Assumptions underlying the definition of Malmquist corrections*

In LB and LS Malmquist corrections are computed by applying Bayes' theorem to obtain $\hat{\omega}$, namely

$$p(\omega_0|\hat{\omega}) = \frac{p(\hat{\omega}|\omega_0)\mathbf{P}(\omega_0)\mathbf{d}\omega_0}{\int p(\hat{\omega}|\omega_0)\mathbf{P}(\omega_0)\mathbf{d}\omega_0} \qquad (9)$$

Both LB and LS assume that the probability density function, $p(\hat{\omega}|\omega_0)$, is a normal distribution, and that $\hat{\omega}$ is unbiased. LB assume the prior distribution, $P(\omega_0)$, of true log distance to correspond to a homogeneous distribution of galaxies. LS, on the other hand, estimate $P(\omega_0)$ by constructing a spline fit to the histogram of log distance *estimates* for the galaxies in the survey, thus in principle taking into account inhomogeneities in the galaxy distribution. Due to the sparseness of surveys, however, it is usually necessary to average the distribution of galaxies over large solid angles, if not all, of the sky. Therefore, the effects of clustering may still go largely unaccounted for (c.f. Newsam et al. 1994).

A more serious problem with the LS Malmquist corrections, however, stems from their use of $\hat{\omega}_{\mathrm{DTF}}$ in Eq. 9. In HS we show that this will result in an *incorrect* Malmquist Correction due to the bias of $\hat{\omega}_{\mathrm{DTF}}$. In general, if $P(\omega_0)$ is constructed from the observed distribution of log distance estimates then one must apply the formula of LS using the ITF estimator. See also the discussion in Teerikorpi (1993) and Feast (1994), where the same conclusion is reached. Malmquist corrections derived from $\hat{\omega}_{\mathrm{DTF}}$ will be valid only when $P(\omega_0)$ is equal to the *intrinsic* distribution of true log distance – an approximation to which one might obtain from, for example, a deeper redshift survey (c.f. Hudson 1994; Dekel 1994). As a special case of this result, note that the homogeneous Malmquist correction (HMC) of LB applied to $\hat{\omega}_{\mathrm{DTF}}$ will be valid provided that the intrinsic distribution of galaxies is homogeneous. This will frequently be a reasonable assumption but is difficult to test. What is certainly clear, however, is that applying the Inhomogeneous Malmquist Correction (IMC) as derived by LS to $\hat{\omega}_{\mathrm{DTF}}$ will be completely inappropriate, since the prior distribution obtained from a histogram of raw log distance estimates will *not* correspond to the intrinsic, but rather to the observed, true log distance distribution.

## 5. POTENT results with Malmquist corrections

In this section we describe the results obtained for a simple velocity field when the estimators and corrections described above are applied to POTENT. For comparison, we also show the effect of uncorrected distance estimates and the problems brought about by using an estimator that does not abide by the requirements of the correction procedure.

Apart from the distance estimators, our analysis follows that of POTENT90 (Dekel et al. 1992), and so the results should be comparable. In particular we use their volume weighted window function (Bertschinger et al. 1990) with a radius of 1200 km s$^{-1}$.

We perform two sets of tests. Both involve Monte Carlo realisations of POTENT velocity field recoveries. The first set of results are for an idealised situation. The underlying velocity field is quiet Hubble flow, and galaxies are drawn randomly from an homogeneous universe with complete sky coverage. Since it is impossible to create a numerical sample with truly infinite depth, galaxies are created homogeneously within a sphere. This sphere is centered around the Milky Way and its radius is such that a galaxy whose absolute magnitude is $M_0 + 3\sigma_m$ is just visible where $\sigma_m$ is the standard deviation of the intrinsic luminousity function.. To generate estimated distances to these galaxies, $M$ and $P$ are sampled from a bivariate normal distribution and subjected to magnitude selection. If the galaxy is unobservable, it is completely discarded. For each observable galaxy, $M$ and $P$ are then used to estimate the distance using all the of the schemes outlined above. We take typical values of the distribution parameters obtained for the $D_n$-$\sigma$ and Tully-Fisher relation that give a log distance error of about 15%. A number of POTENT realisations for each method are calculated and the average used to show the biases. This test is designed to show the effect of the various distances estimation and correction techniques in what is, in some sense, a 'best case' situation. The use of quiet Hubble flow will ensure that no sampling gradient biases are introduced and the homogeneous universe is ideal for testing the assumptions of the Malmquist corrections.

The second set of tests are for a more realistic situation. Quiet Hubble flow is replaced with a more complex peculiar velocity field involving a large void and an attractor region. Also, the galaxies are positioned much more realistically.

This realistic distribution is generated as follows. We position a galaxy at each point in the combined data of Mathewson et al. (1992) and Burstein et al. (1987) (see Fig. 1). For simplicity, we neglect the improvement in the distance estimates that can be found for clusters by combining estimates from all the galaxies in the cluster. However, in order to avoid "finger of God" type effects, one galaxy is chosen at random from each cluster. Again, $M$ and $P$ are drawn from the same bivariate normal and the same selection applied. However, if the galaxy is found to be unobservable, $M$ and $P$ are regenerated and the process repeated until the galaxy is 'observed'.

in the POTENT realisations (see Dekel et al. 1992) and enables easier and more accurate comparison. Figure 2 shows a slice through this smoothed field.

*5.1. Monte Carlo* POTENT *velocity field recoveries*

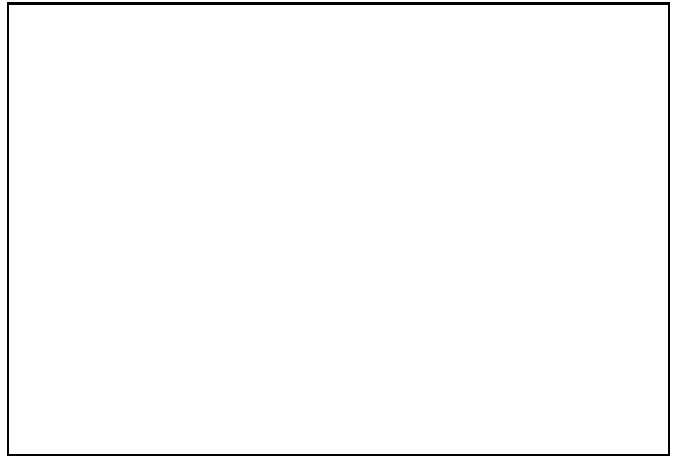

**Fig. 3.** Bias as a function of distance for POTENT recoveries of quiet Hubble flow with galaxies drawn from an homogeneous universe. The DTF estimator has been used on its own, with an homogeneous Malmquist correction and with an inhomogeneous correction. The distance estimator has an error in log distance of about 15%.

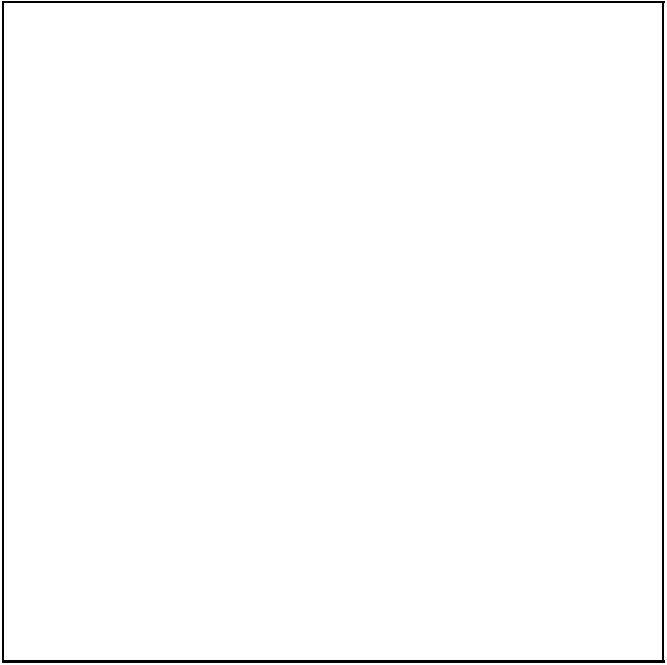

**Fig. 2.** A slice through the center of the velocity field used. The void and attractor regions are modelled by deriving the field from a potential which contains a large gaussian well and a wider gaussian peak. Note that this graph actually shows the field smoothed with a gaussian window of 1200 km s$^{-1}$ radius to enable direct comparison to the POTENT results.

The field we have chosen to use for these tests is formed by creating a potential consisting of two large spherically symmetric gaussian fluctuations, one positive, centered around (4000, 0), and one negative at (-2000, 2000). The velocities are then constructed as in Eq. (3) and these velocities used to assign redshifts to the galaxies. When comparisons are made to POTENT results, it is necessary to smooth the field with a gaussian window of 1200 km s$^{-1}$. This is the radius of the window used

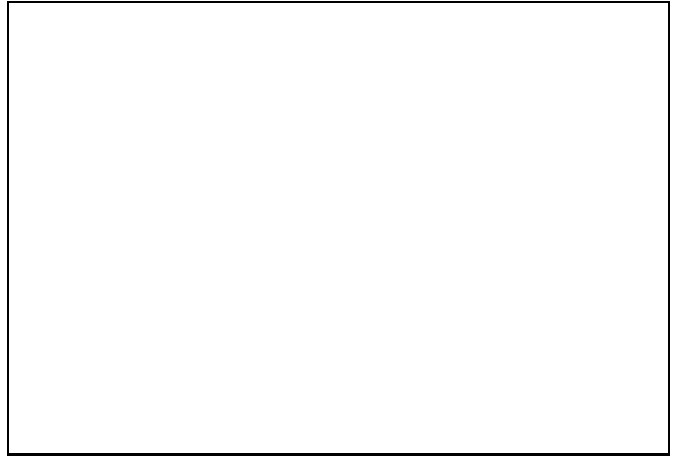

**Fig. 4.** Bias as a function of distance for POTENT recoveries as in Fig. 3. This time, the ITF estimator has been used, both 'raw' and with the corrections.

The results of the first, simpler test are shown in Figures 3 and 4. Since we are dealing with such a simplified galaxy sample and velocity field, all the bias is in the radial direction and is the same along all radial lines. These graphs show this bias as a function of distance. The need for some form of correction is clear, particularly for the DTF estimator, as are the dangers of an inappropriate correction – for example, the homogeneous Malmquist correction applied to the ITF estimator which is far worse than using the uncorrected distances. Slightly surprising is the good recovery from the inhomogeneous correction when applied to DTF type estimates. However, we know that the biased, non-gaussian nature of the distance estimator violates some basic assumptions of the correction and such a complex form of bias from such a simple sample distribution does not augur well for more realistic galaxy surveys. Figure 5 highlights this. Here the $M$ and $P$ distribution parameters have been changed to give only 10% log distance errors. Suddenly the inhomogeneously corrected recovery is dramatically worsened showing that the good results with 15% errors were just a lucky coincidence.

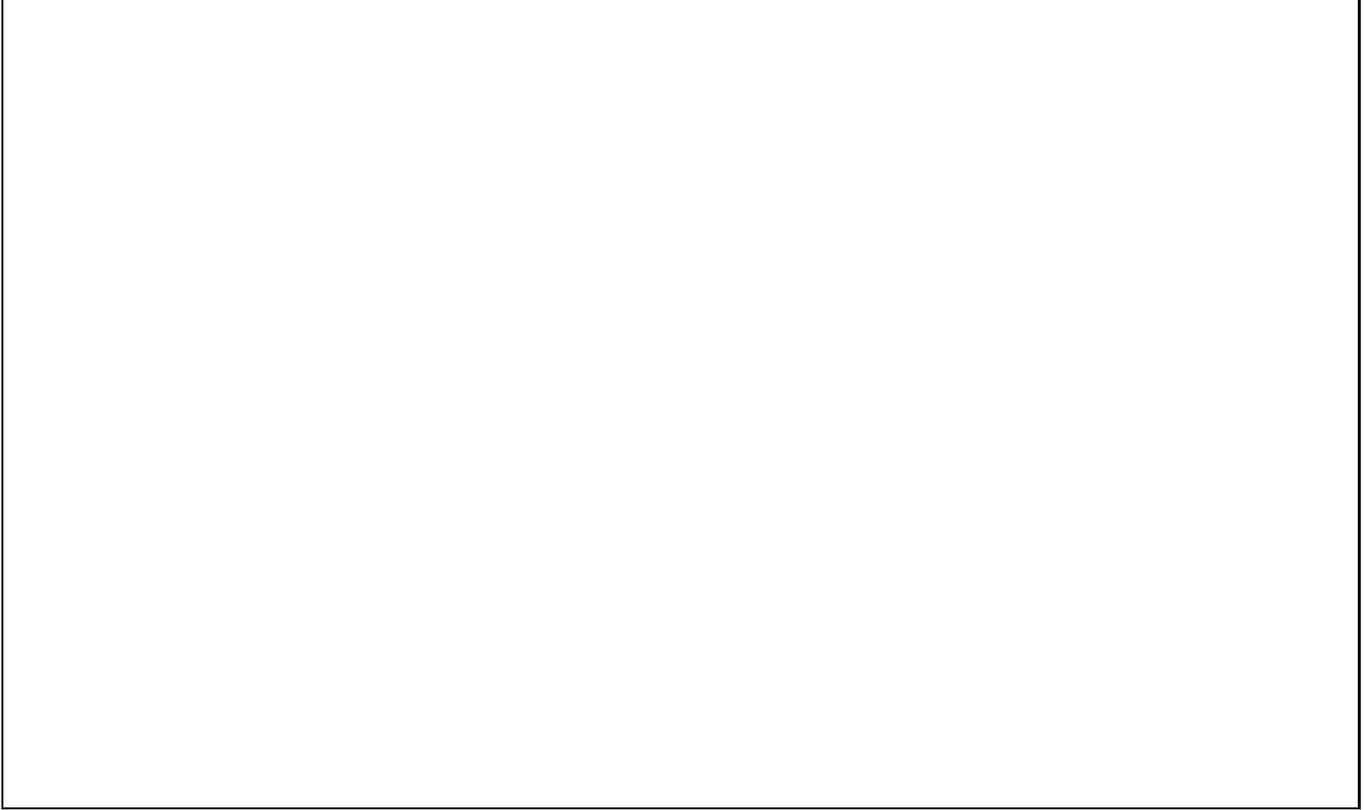

**Fig. 1.** An Aitoff projection of the galaxies used as the sample for the more realistic Monte Carlo realisations. Coordinates are galactic longitude and latitude. In particular, note the large void centered around longitude 90° where the Mathewson sample is particularly sparse.

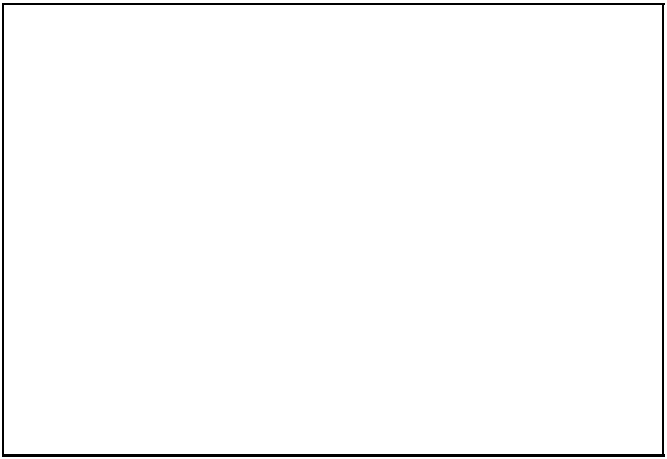

**Fig. 5.** Bias as a function of distance for POTENT recoveries as in Fig. 3. Again, DTF estimators are used, but this time they have only 10% log distance errors.

We have found that such cancellations are quite common in our Monte Carlo tests. Because of the many sources of systematic errors in POTENT, circumstances are bound to occur where two or more biases approximately cancel out. It is only by rigorously performing tests for a large variety of distances estimators and errors, galaxy distributions, velocity fields and so on, that such coincidences can be recognised and the true advantages and problems of any method brought to light. Too few tests may well make inappropriate methods seem attractive. Considerable care needs to be taken to avoid falling into traps of this sort.

The best recoveries, therefore, are as expected. Since we are drawing from an homogeneous universe, the homogeneous correction is indeed effective when used with the direct regression line estimator and the recovery from an ITF type estimator with an inhomogeneous correction is equally good (Fig. 4).

So, the theory holds out well for this situation and the corrections, when properly applied, seem to be adequate. However, we need to consider the effect of inhomogeneities in the universe, incomplete samples and more complex velocity fields.

We perform the second, more realistic test on the same distance estimation techniques described above. The results of these tests are shown in Fig. 6. For clarity, in Fig. 7 the actual smoothed field has been subtracted from the Monte Carlo recoveries to give just the bias. As expected, the inhomogeneous correction provides an improvement over the 'raw' ITF recovery, although not a particularly substantial one. However, again, the inhomogeneously corrected DTF is quite good and surprisingly, the 'raw' DTF recovery is the best of the lot and the homogeneous correction makes it far worse. Since we know that in the idealised case exactly the opposite is true, the DTF recovery must be another "lucky" coincidence. These coincidences come about because of the complexity of the biases in POTENT: Malmquist type biases come out of the smoothing, the distance estimates themselves can be intrinsically biased, inhomogeneous galaxy distributions lead to pollution or sampling gradient biases and so on. The interaction between all the errors is complicated and unpredictable. Although the chances of a fortunate cancellation exist, they cannot be relied upon and the chance of an unrecognised bias getting through the correction procedure is high. One such bias can be seen in all four recoveries. In the lower right hand quadrant of the figures the is a large inward bias. This is caused by a large gap in the coverage of the Mathewson et al. (1992) data set and it appears to shift the void to considerable greater distances in all cases. In fact, the situation is not quite as bad as it may seem as areas such as these have relatively large random errors and are usually discarded from the recovery before any analysis is done, but the danger remains.

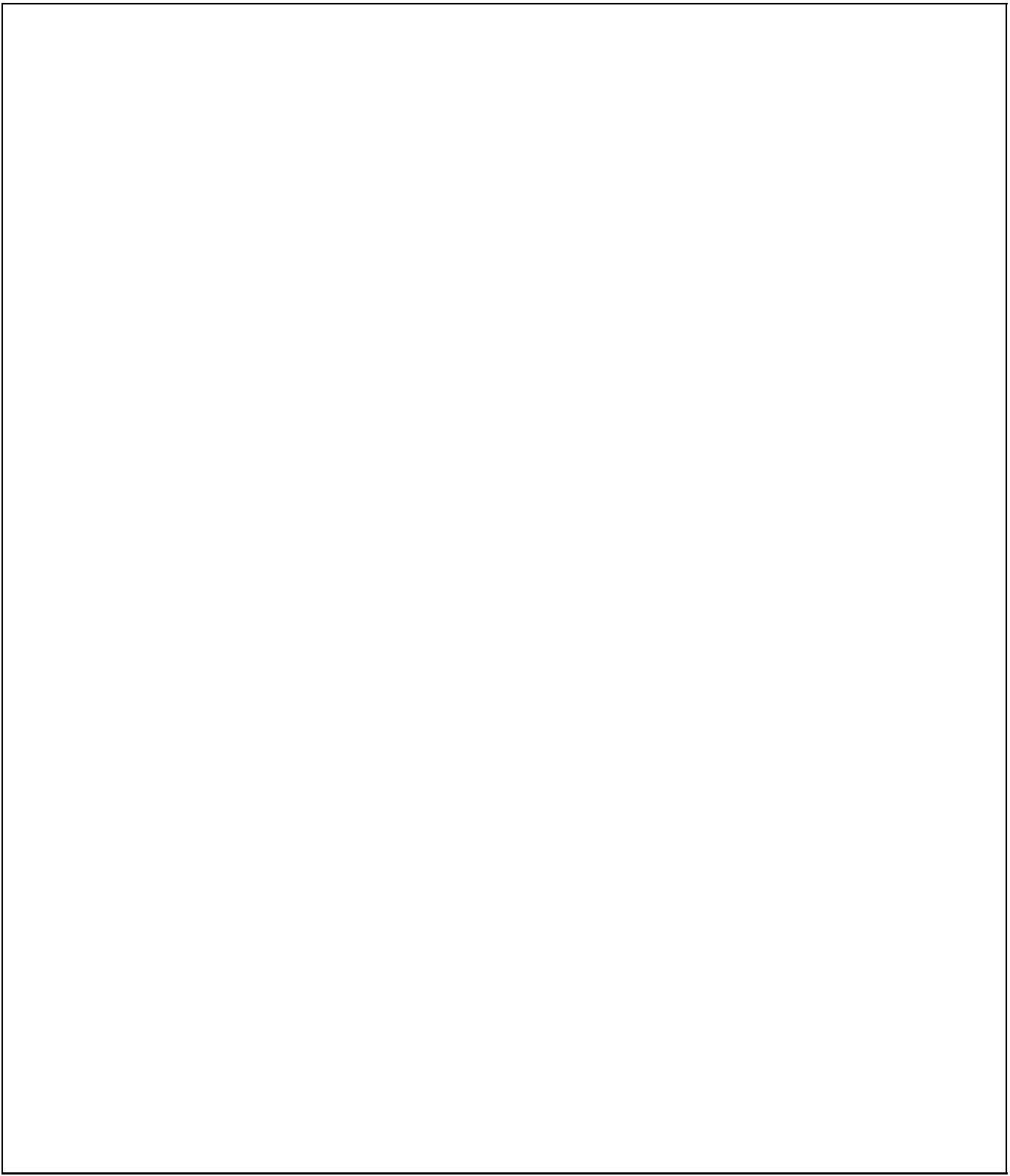

**Fig. 6.** Monte Carlo POTENT recoveries of test velocity field shown in Fig. 2 using four different methods of distance estimation. In graphs (a) and (b), 'raw' DTF and ITF distance estimates are used. Graphs (c) and (d) have homogeneous corrections applied to the two 'raw' estimators and in (e) and (f), the inhomogeneous corrections have been used.

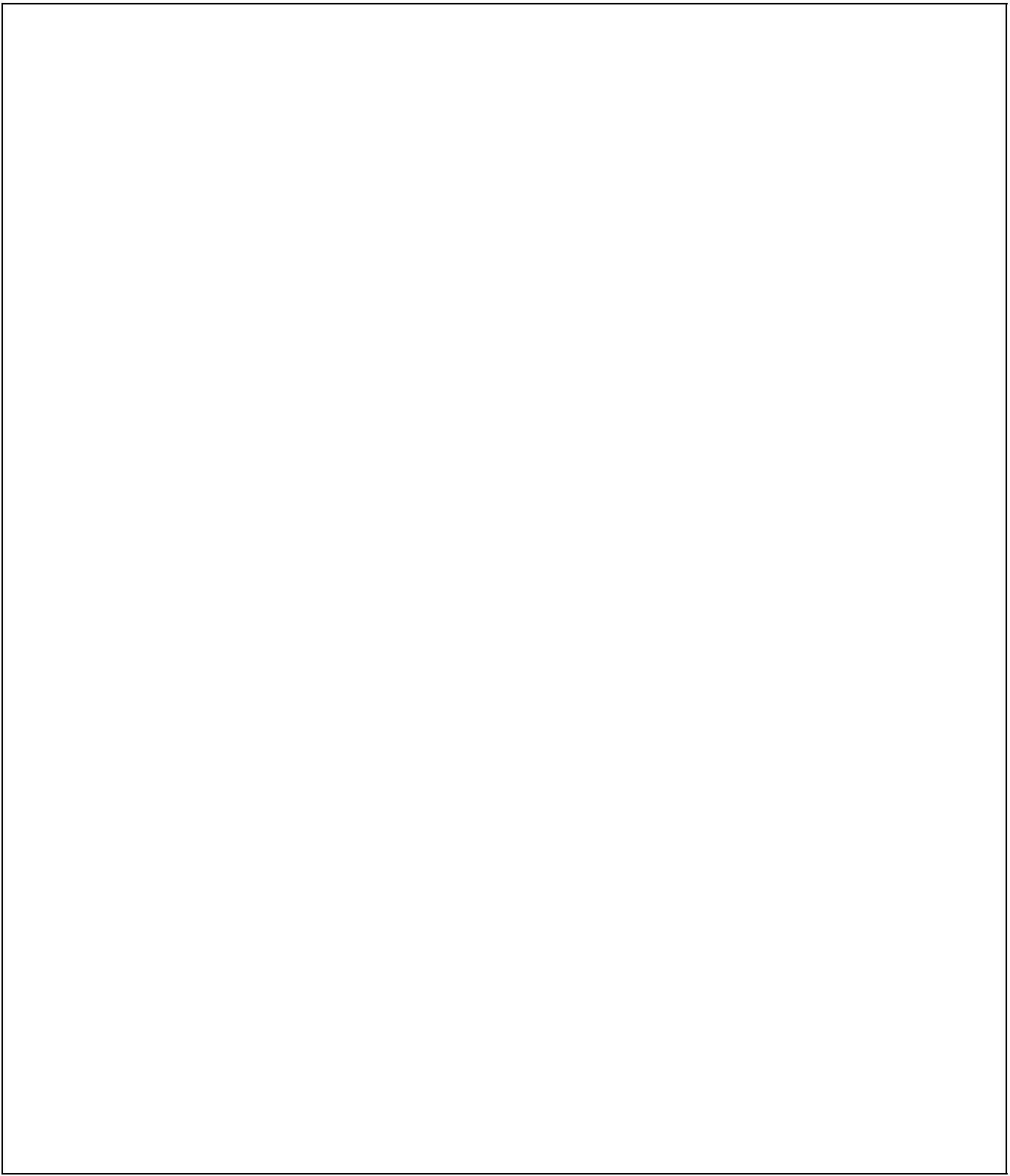

**Fig. 7.** Biases of the four recoveries shown in Fig. 6. The actual smoothed velocity field has been subtracted from the Monte Carlo POTENT recoveries.

The problems with the homogeneous correction with both 'raw' estimators are twofold. Firstly, the assumption of an homogeneous sample implied when ITF estimates are used is clearly not valid (see Fig. 1) and the universe is probably not homogeneous enough on these scales to enable us to use the DTF. However, these effects are not sufficient to account for the large spurious inflows seen here. These are due to the modelling of the selection effects in the Monte Carlo simulations. Since we have a combination of two data sets and they are being used solely to provide realistic distributions rather than actual data to be processed, the choice of magnitude cut-off and the sharpness of that cut-off will not exactly match the real data. Therefore, even if the real universe were homogeneous on these scales, the universe of our fake surveys will not be, since they are forced to reproduce the same sample distribution with different selection criteria. This is, therefore, an artifact of the Monte Carlo technique rather than the Malmquist corrections used, but the magnitude of the effect in POTENT recoveries should be taken as a warning both for the design of Monte Carlo type tests and for the use of the homogeneous correction when the homogeneity requirement is suspect.

Other features are apparent in all the recoveries. In particular, there is a large spurious outflow from a region centered loosely around (8000, -3000). This is a sampling gradient bias brought about by the relative sparseness of galaxies in this region, as can be seen in Fig. 1. Here the large void caused by the incomplete sky coverage of the Mathewson et al. (1992) sample is clear.

Given these problems we feel that an alternative approach is called for. We will here outline one such approach and present some numerical tests on realistic data sets to compare it to the methods described so far.

## 6. Iterative Monte Carlo corrections

Our standpoint is to try and simplify the problem by moving away from the details of how all the different biases actually come about, and treat them all jointly as a systematic error that needs to be calculated and corrected. We therefore wish to use POTENT itself to find these errors, for example using Monte Carlo recoveries similar to the ones shown in Fig. 7, and then simply subtract them from the actual recovery.

To do this, we divide the systematic errors into two classes; those that do not depend on the underlying velocity field and those that do. For example, the Malmquist bias is independent of the velocity field, but sampling gradient biases have a strong dependence. The former can be dealt with under the assumption of any velocity field (we take quiet Hubble flow for simplicity), but the latter requires some means of iterating towards a solution. Before we get to this, however, we need to carefully choose a Monte Carlo method.

### 6.1. Monte Carlo procedure

We need a procedure that uses the distribution of galaxies given by the data set but can impose some chosen velocity field onto them. We also need to be sure that the details of the distance estimator and selection of galaxies are consistent with those that went into forming the original sample. Finally, we need to be certain that no systematic errors are introduced by the procedure itself, so that the results can be confidently treated as a correction.

We therefore use the procedure outlined below.

1. Create a mock universe of observable galaxies by assigning some definite position $\vec{r}_f$ to each galaxy in a catalogue observed at some $\hat{\vec{r}}$. (For simplicity, this is done by $\vec{r}_f = \hat{\vec{r}}$ for all galaxies. However, it may be better in some cases to use the redshift as a position for more distant galaxies).
2. Imposing some velocity field $\vec{v}_f(\vec{r}_f)$, assign a redshift $z_f$ to each galaxy such that

$$cz_f = u_f - H_0 r_f \qquad (10)$$

3. For the example of an estimator based on two observables $M$ and $P$, assign an $M$ and a $P$ to each galaxy. These will be randomly sampled from a distribution whose parameters are estimated from the original data set.
4. Impose selection on each galaxy and, if unobservable, go back to step 3.
5. Get estimates of distances to galaxies $\hat{\mathbf{r}}_f$.
6. Use $\hat{\mathbf{r}}_f$ and $z_f$ in POTENT to get a recovered velocity field.
7. Repeat steps 3 to 6 a suitable number of times and average the resulting velocity field.

This will result in a good approximation to the systematic errors introduced by POTENT when acting on a particular velocity field with a particular galaxy catalogue. The iterative process involves modifying the imposed velocity field $\vec{v}_f(\vec{r}_f)$ at each stage and, hopefully, converging it towards the underlying smoothed field.

### 6.2. Application of iterative corrections

The basis of the iteration is to use the monte carlo results of some 'guess' velocity field to give a next guess that is a little bit closer to the underlying $\vec{v}$.

We shall call the first guess field $\vec{v}_f^{(0)}$. We then apply our monte carlo procedure to obtain an estimate of the systematic errors $\vec{b}_r(\vec{r})$ this field produces. For simplicity of notation, we will call this monte carlo velocity field $\bar{\vec{v}}_f^{(0)}$ where the bar denotes the average including all biases. Thus

The next iteration could then be found by removing these biases from the single POTENT recovery of the original catalogue. However, this recovery is noisy and using it as the basis of an underlying velocity field in the iterative scheme will try and force the method to fit the noise. In order to avoid this, it is necessary to perform a monte carlo on the 'raw' data by scattering the distances as above, but using the actual redshifts from the catalogue to define the imposed velocity field. This will give us an averaged field $\bar{\vec{v}}_{\mathrm{raw}}(\vec{r})$ which well approximates the combination of systematic errors from POTENT and the real smoothed velocity field.

We then obtain the next iteration velocity field:

$$\vec{v}_{\mathrm{f}}^{(1)} = \bar{\vec{v}}_{\mathrm{raw}} - \vec{b}_{\mathrm{f}}^{(0)} \qquad (12)$$

The monte carlo procedure can then be repeated so that, in general, to obtain the velocity field for iteration n,

$$\vec{v}_{\mathrm{f}}^{(n)} = \bar{\vec{v}}_{\mathrm{raw}} - \vec{b}_{\mathrm{f}}^{(n-1)} \qquad (13)$$

6.2.1. Convergence criteria

Convergence will have occurred when, within some tolerance,

$$\vec{v}_{\mathrm{f}}^{(n)} \approx \bar{\vec{v}}_{\mathrm{raw}} \qquad (14)$$

However, exactly how to define the comparison of the two fields is not simple. The simplest approach would be to do a point by point comparison of the two grids and average all the values obtained. Unfortunately, this approach has two problems. Firstly, some areas of the POTENT recovery will be extremely noisy (particularly at large distances) and to get the residual noise in the monte carlos down to really acceptable levels would be prohibitively time consuming. In addition, the large gaps in galaxy surveys, whether they are due to actual voids or incomplete sky coverage, give areas where there is little or no information about the velocity field. Since the correction method acts by adjusting the redshifts of the galaxies it is using, where these galaxies are sparse very little effect can be achieved. However, if the convergence criteria gives too much weight to these regions, they may prevent the tolerance level being reached and the corrections will 'over-shoot'. In practice, areas which are subject to these problems are discarded before any analysis of the POTENT results is performed, so we would wish to minimise their effect on the convergence criteria without making any ad hoc decisions about 'good' and 'bad' areas of the field.

We do this in two simple ways. Firstly, instead of comparing at grid points, we interpolate the fields to the 'fixed' positions of the galaxies and perform the comparison there. In this way, we put most weight on densely sampled regions where the recovery is likely to be most useful. Also, the variance of the $\bar{\vec{v}}_{\mathrm{raw}}$ recovery is calculated and used to weight the comparison at every point, again making the correction best where it is needed. This weighted average is used as the value of the current iteration, and when it drops below some tolerance (usually in the tens of km s$^{-1}$), the process is stopped.

7. Results using iterative correction on POTENT

The tests in this section are for direct comparison to the second, more realistic set of results in Sect. 5. Therefore, the underlying velocity field is the same as are the distribution of galaxies and the parameters used to create the data for the distance estimates. Although the method can, in theory, be used with any number of different distance estimators, we have chosen to use an ITF estimator. There is no overwhelming reason why this is the best estimator to use, but since it is unbiased, it will not introduce any systematic errors of its own, thereby perhaps speeding up convergence. In addition, its gaussian nature makes it easy to perform analysis on.

For the tests shown in this section, we used a convergence tolerance of 50 km s$^{-1}$ where the convergence criteria is as described in Sect. 6.2.1. In order to ensure that residual noise in the Monte Carlo averages was not significant at this level, each Monte Carlo recovery used 200 POTENT realisations of the velocity field.

All that remains to be decided, therefore, is the choice of the initial guess field $\vec{v}_{\mathrm{f}}^{(0)}(\vec{r})$. We choose quiet Hubble flow partly because it has no features that might pose unreasonable constraints on the recovery method, and also, since all peculiar velocities are zero, there are no sampling gradient biases in the recovery of quiet Hubble flow. This means that the bias of the initial recovery $\vec{b}_{\mathrm{f}}^{(0)}(\vec{r})$ is due solely to the effects of smoothing over the galaxy distribution distorted by distance errors. Therefore, this bias can be used as a form of Malmquist correction that takes into account *all* the inhomogeneities in the sample and makes no assumptions about the form of the distance estimator. If this is all that is required, the process can be stopped there.

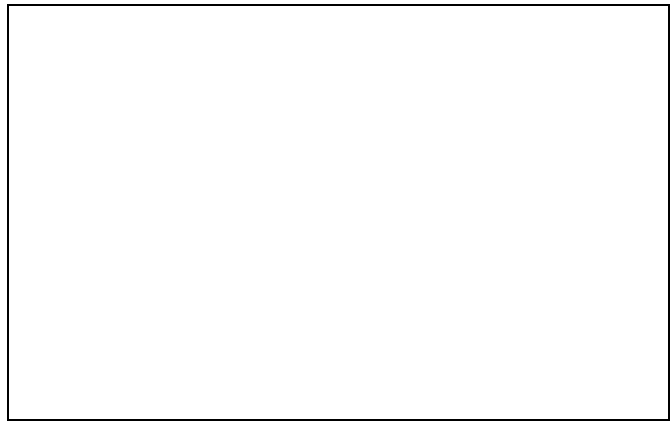

**Fig. 9.** A comparison of the various methods. For the purposes of the comparison, the Monte Carlo recoveries shown in previous figures are interpolated on to cubic grids with grid spacing of 500 km s$^{-1}$. The shaded bars show the average errors of a simple point by point comparison of these interpolated fields with the actual smoothed velocity field. The hollow bars are the same comparison with each point weighted by the variance of the POTENT recovery at that point.

The usual slice through the final velocity field is shown in Fig. 8 together with its bias. The recovery is considerably better than any of the previous results. Just how much better can be seen in Fig. 9. Here, the various averaged recovered velocity fields have been interpolated onto a cubic grid truncated to the 8000 km s$^{-1}$ sphere. The grid spacing is chosen to be

**Fig. 8.** Average POTENT recovery of field given in Fig. 2 using Monte Carlo iterative correction method. Graph (a) is from the final corrected recovery and (b) is its bias.

500 km s$^{-1}$ for comparison with the POTENT$_{90}$ of Dekel et al. (1992). Then, we simply compare the velocity at each grid point with the actual smoothed velocity field and average all the errors. The weighted comparison uses the variance of the POTENT recovery to weight each point. With both comparisons the Monte Carlo correction is better, considerably so without the weighting.

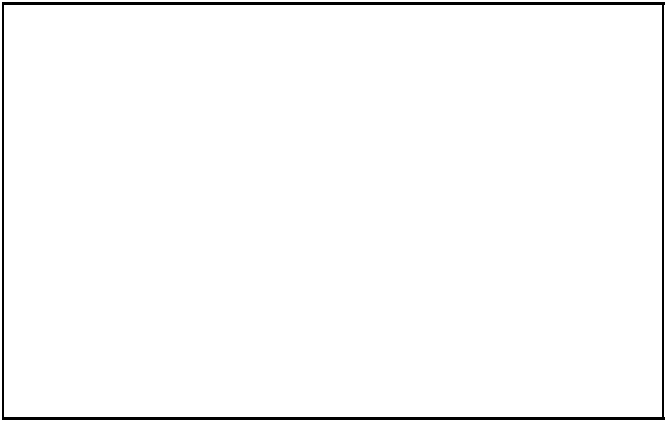

**Fig. 10.** The average biases of the Monte Carlo correction technique at each iteration. As with Fig. 9, the shaded bars show the average of a point by point comparison of the recovered average field with the smoothed real field on a cubic grid. For the hollow bars, the comparison was weighted at each point by the variance of the POTENT recovery. The thick line shows the level of the convergence criteria value as described in Sect. 6.2.1.

In Fig. 10, we perform the same comparisons with each iteration of the Monte Carlo correction. The fourth iteration is the one we used since this is the one where the convergence level dropped below 50 km s$^{-1}$ (shown by the thick line in the figure). However, it is clear that the best iteration is the second one, if only by a small margin. This is caused in part by corrections in the sparse regions over-shooting since the limited number of galaxies gives the correction very little to work with and so it continually tries to remove the same bias. But the small residual bias in the final iteration (Fig. 8) even in the very sparse region mention earlier (Fig. 1) shows that this is a relatively small effect, and something else is needed to account for the seeming divergence. To understand what, we need to recall that the convergence level was calculated using a comparison involving $\vec{v}_{\rm raw}(\vec{r})$. This is only an estimate of the combination of systematic errors from POTENT and the real smoothed velocity field. The fact that the method is as successful as it is is a testament to the fact that it is a fairly good estimate, but down at the levels of tens of km s$^{-1}$ which we are considering, this assumption must fall down. Therefore, if convergence levels this low are really required, then some means of improving this estimate will need to be found. However, most of the work is clearly done in the first two or three iterations, so convergence levels of about 100 km s$^{-1}$ would really be more useful, and would considerable reduce the computational overhead.

## 8. Conclusion

We have seen, therefore, that errors in distance estimates give rise to large and complicated systematic errors in POTENT even when the estimators themselves are unbiased. Particularly significant are those caused by the grouping together of galaxies in the smoothing windows where we wish to group together galaxies according to their actual position in space, but can only group by estimated position. The biases produced in this way depend both on the characteristics of the distance estimator and on the actual distribution of galaxies. One of the so-called Malmquist corrections are frequently used to compensate for them.

However, these corrections must be very carefully applied. Both the homogeneous correction of LB and the general, or form of the distance estimator and distribution of galaxies and unless these are valid for a particular data set, the corrections will simply introduce new biases. In particular, when using the common Tully-Fisher or $D_n-\sigma$ type relations, it is very important to use the appropriate regression line to relate the two observables. Therefore, if a homogeneous universe is assumed, the homogeneous Malmquist correction must be applied to direct regression line estimates, whereas the inhomogeneous correction requires the unbiased characteristics of distance estimates drawn from the inverse regression.

The tests in this paper show that, for simple cases, the Malmquist corrections do provide considerable bias reduction when correctly applied. However, the bias removal is far from perfect since the use of the corrections in this context is not fully justified for the large window functions needed by POTENT.

The situation worsens considerably for more realistic tests with complicated galaxy distributions and inhomogeneous velocity fields. Here more complex Malmquist-like biases are introduced in the transverse direction where neither correction can deal with them, and some of the other biases in POTENT, especially sampling gradient biases, become significant. These extra biases, as well as degrading the quality of the recoveries, also lead to 'lucky' and 'unlucky' combinations of biases which confuse analysis of the tests. We therefore propose a new approach to bias removal in POTENT.

This new approach uses the results of Monte Carlo tests to iteratively improve on some initial guess velocity field. Because the Monte Carlos use the distribution and parameters of the survey data directly, the corrections are based in the interaction between the data and POTENT and should, therefore, produce tailored corrections.

There are many possible problems to the method. It may not converge, or worse may converge to the wrong field, because of residual noise in the Monte carlo recoveries. This can be overcome by increasing the number of POTENT realisations per iteration, but this is expensive in computer time. There is also a risk of overcorrecting regions very sparse in galaxies. However, these areas will probably be discarded before the recoveries are used due to the large random errors in POTENT in sparse neighbourhoods. Also, since the tests show that most of the correction comes in the first iterations, less stringent convergence criteria would help to alleviate this problem as well as reducing the computational expense without significant loss in the bias removal.

The initial results presented in this paper and other simpler tests we have performed show that the iterative corrections perform noticeably better than any of the other methods considered. Even were this not the case, and the bias levels were similar to the best of the other methods, the technique has another advantage. Because corrections deal with bias directly from the interaction between POTENT and data, the distance estimator and parameters of the window function in POTENT can be chosen to minimise the noise of the recovery rather than the bias and leave the iterative Monte Carlos to do the rest.

The method now needs to be applied to a careful combination of the best data currently available and compared to results currently in use. Whatever the outcome, it offers a new avenue for bias removal for a variety of techniques, not just POTENT, and holds much promise for future application and improvement.

## 9. Acknowledgements

A.N. was funded during this research by SERC Postgraduate Studentship 91306795. M.A.H. was funded by an SERC Postdoctoral Fellowship.